# A hybrid LBM-DEM framework with an improved immersed moving boundary method for modelling complex particle-liquid flows involving adhesive particles

Wenwei Liu and Chuan-Yu Wu[*]

Department of Chemical and Process Engineering, University of Surrey,

Guildford, GU2 7XH, UK

**Abstract**

This paper presents an improved immersed moving boundary model (IBM) for solving complex fluid-particle interactions in a coupled lattice Boltzmann method (LBM) and an adhesive discrete element method (DEM), using the "partially saturated cell" scheme. It is shown that the existing scheme does not well address the contribution of each solid particle to the fluid when multiple particles intersect with the same lattice cell. This issue is completely addressed by modifying the weighting function in the partially covered cells in the present study. Furthermore, a fast linear approximation method with high efficiency and good accuracy is applied to calculate the partially intersected volume between a particle and a lattice cell. Verified with several benchmark cases, the developed hybrid IBM-LBM-DEM numerical framework is capable of describing the flow field between dense particles with a relatively low grid resolution in more details, as well as effectively capturing the adhesive mechanics between microspheres.

[*]**Corresponding author:** Prof. Chuan-Yu Wu, **Email:** c.y.wu@surrey.ac.uk

# 1. Introduction

The transport of solid particles with fluids ubiquitously exists in a variety of scientific issues and engineering applications. In these problems, the fluid-particle interactions are of great significance to well describe the mechanics of individual particles, which are crucial to understanding the behaviour of the system. Because of its complex nature, it is still challenging to accurately characterise the fluid-particle hydrodynamic interactions. Nevertheless, a few numerical techniques are developed for analysing the complex fluid-particle multiphase flows. For example, the computational fluid dynamics (CFD) coupled with the discrete element method (DEM) is the most widely used numerical approach [1-7]. In addition, the lattice Boltzmann method (LBM) becomes a promising alternative approach for solving fluid flows [8-13] in the recent two decades, which shows significant flexibility in handling complex boundary conditions and provides a high space-time resolution. In LBM, the fluid domain is generally discretised with a regular orthogonal grid, which is called lattice and is similar to the mesh grid in CFD. The fluid in each lattice is represented by packets of fictitious 'particles', of which the motions follow the lattice Boltzmann equation. During each computational time step, the fluid flow in a lattice is updated only with its local information, which makes it naturally suitable for parallel computing.

For the fluid-particle interactions in LBM, several coupling techniques were proposed, including the bounce-back scheme (BB) [14-21], and the immersed boundary method (IBM) [22-27]. The BB scheme uses a discretised lattice representation of solid particles, which is the most popular one due to its physical simplicity and robustness to describe any particle shape. Ladd [14-15]12 introduced a modified BB scheme, assuming that the bounce back occurs at the half-way between the solid lattice nodes and fluid lattice nodes. The basic idea of this modified BB scheme is that the non-equilibrium part of the distribution on a boundary link remains unchanged during bounce back. However, the modified BB scheme is known to be of second-order accuracy only when the solid surface is located at just half lattice spacing from the boundary node. The accuracy for modelling a curved surface, which is approximated by zig-zag staircases, is of first order only, and large fluctuations of the hydrodynamic interactions are also observed [21]. To capture the actual shape of the solid surface and maintain the second-order accuracy, several modified bounce back schemes based on the spatial interpolation were also proposed [16-21]. Bouzidi et al. [16]16 considered the relative location of the boundary node with a location parameter $q=|\mathbf{x}_f - \mathbf{x}_w|/|\mathbf{x}_f - \mathbf{x}_b|$, where $\mathbf{x}_b$, $\mathbf{x}_w$, and $\mathbf{x}_f$ represent the locations of the solid node, the solid boundary node and the fluid boundary node, respectively. Two different interpolation expressions are derived for $q>0.5$ and $q<0.5$, respectively, either with linear interpolation or quadratic interpolation. Filippova and Hänel [17] proposed an interpolation scheme based on a weighting parameter and a fictitious equilibrium distribution, both of which depend on the relative location parameter $q$ as used in the Bouzidi's scheme. Hence, this interpolation scheme also has two different expressions with respect to the value of $q$. Further improvements were made by Mei et al. [18-19] to increase the numerical stability. Yu et al. [20] proposed a double interpolation scheme, where separate treatments of the interpolation for $q>0.5$ and $q<0.5$ are replaced by three steps in the realisation of the boundary treatment. As a result, all values of $q$ are handled with the same expression. All these interpolated BB schemes not only maintain the second-order accuracy of LBM, but also generate much more accurate and smoother results for the hydrodynamic force and torque evaluation.

The immersed boundary method (IBM) proposed by Peskin [22-23] was widely implemented in the CFD-DEM models [28], and further coupled in LBM by Feng and Michaelides [24]. In this method, the fluid and the particles are represented with fixed Eulerian mesh grids and the moving Lagrangian nodes, respectively. The particles are assumed to be deformable with a large stiffness. The effects of the immersed particle boundary on the fluid are first modelled by restoring forces, which tend to keep the particle to its original shape, on the Lagrangian grid based on the no-slip boundary condition. Then the restoring forces are distributed to their surrounding Eulerian grids and considered as the external force terms in the governing equations. The drawback of the IBM lies in the introduction of additional free parameters. The values for the spring stiffness and the damping constant have to be determined in a problem-dependent fashion. Moreover, the time step may be severely restricted, as the characteristic time scales of the oscillations of the spring-damper systems need to be resolved [25]. IBM was further improved by Uhlmann [25] and Breugem [26] in order to introduce the force term that is not determined by any feed-back mechanism as well as suppress the oscillations caused by the fixed grid. Such an improved IBM was coupled with LBM and applied in the simulation of particle-laden turbulent flow involving dense particle suspensions [29-30].

The "partially solid scheme" developed by Noble and Torczynski [27] as a modified IBM was widely applied in the numerical study of dense fluid-particle flows [31-37]. An additional collision operator is introduced to the conventional lattice Boltzmann equation, which is modified with a weighting function that depends on the solid fraction in the local lattice cell.

As the weighting function varies between 0 and 1, the lattice Boltzmann equation for pure fluid nodes as well as the completely bounce back for pure solid nodes can be recovered. Therefore, this IBM maintains the local simplicity of handling the lattice Boltzmann equation and is much easier to implement. The only challenge is the accurate estimation of the solid fraction in each lattice cell, for which some issues were still not properly addressed. In particular, when multiple particles intersect with the same lattice cell, the contributions to the total weighting function from each intersecting particle are not considered in a self-consistent way, which can often cause numerical divergence. Furthermore, the hydrodynamic interactions on each solid particle in the same cell should also be weighted based on its corresponding solid fraction contribution. Such discrepancies are generally negligible in dilute particle-fluid flows, as interparticle contacts barely occur. Nevertheless, for the dense multiphase flow where numerous close contacts between particles are developed, it could lead to severe numerical problem, such as divergence and irrational particle collisions.

Therefore, in this paper, an enhanced hybrid numerical framework is introduced for modelling particle-fluid flows with adhesive particles based on a coupled single-relaxation-time LBM and DEM. An improved IBM method is developed for solving the fluid-particle interactions, where the solid fraction is determined with a new linear approximation method of high efficiency and accuracy. In addition, the Johnson-Kendall-Roberts (JKR) adhesive contact mechanics is adopted in the DEM to describe the interparticle normal force for adhesive particles. To the best of our knowledge, there were very limited LBM-DEM studies on two-phase flows with adhesive particles, despite the ubiquity of their applications in almost all areas of engineering, biology, agriculture and physical sciences [38-46]. The challenge lies in coupling adhesion, elastic contact forces and frictional forces in the short-range particle-particle interaction zone as well as the coupling with the fluid forces (e.g. buoyancy, drag and lubrication) across both the long-range length scale and the time scale. In the current work, an implicit solution is employed to obtain the particle-particle adhesive normal force. The paper is organised as follows. The numerical model is introduced in Section 2 in details, where a brief review of the LBM, the improved IBM, the fast linear approximation method to determine the solid fraction, as well as the DEM for adhesive particles, are described. Section 3 presents the model validations, and an application of the LBM-DEM to the random packings of microspheres is discussed in Section 4. Conclusions are drawn in Section 5.

# 2. Numerical model

## 2.1 Lattice Boltzmann method (LBM)

In LBM, the fluid domain is discretised into non-overlapping orthogonal grid, which is called lattice. The fluid in every lattice is assumed to be represented by fictitious moving particles, which are described by a statistical density distribution functions $f_i(\mathbf{x},t)$. The fluid 'particles' first undergo a local relaxation process in the lattice, also known as the collision step, and then propagate across lattices, which is called the streaming step. During the propagation, the fluid 'particles' are only allowed to move in certain directions with certain discrete speeds. So far there are dozens of lattice speed models proposed in the literature, such as D2Q7, D2Q9, D3Q15, D3Q19, D3Q27 and so on, where the first number denotes the dimension and the second number represents the number of speed directions. In this work, the widely used 3D discretisation schemes D3Q19 model is adopted, as displayed in Fig. 1, where the fluid 'particles' in each lattice move to its 26 immediate neighbours with 18 different velocities, $\mathbf{e}_i$ ($i$=1~18).

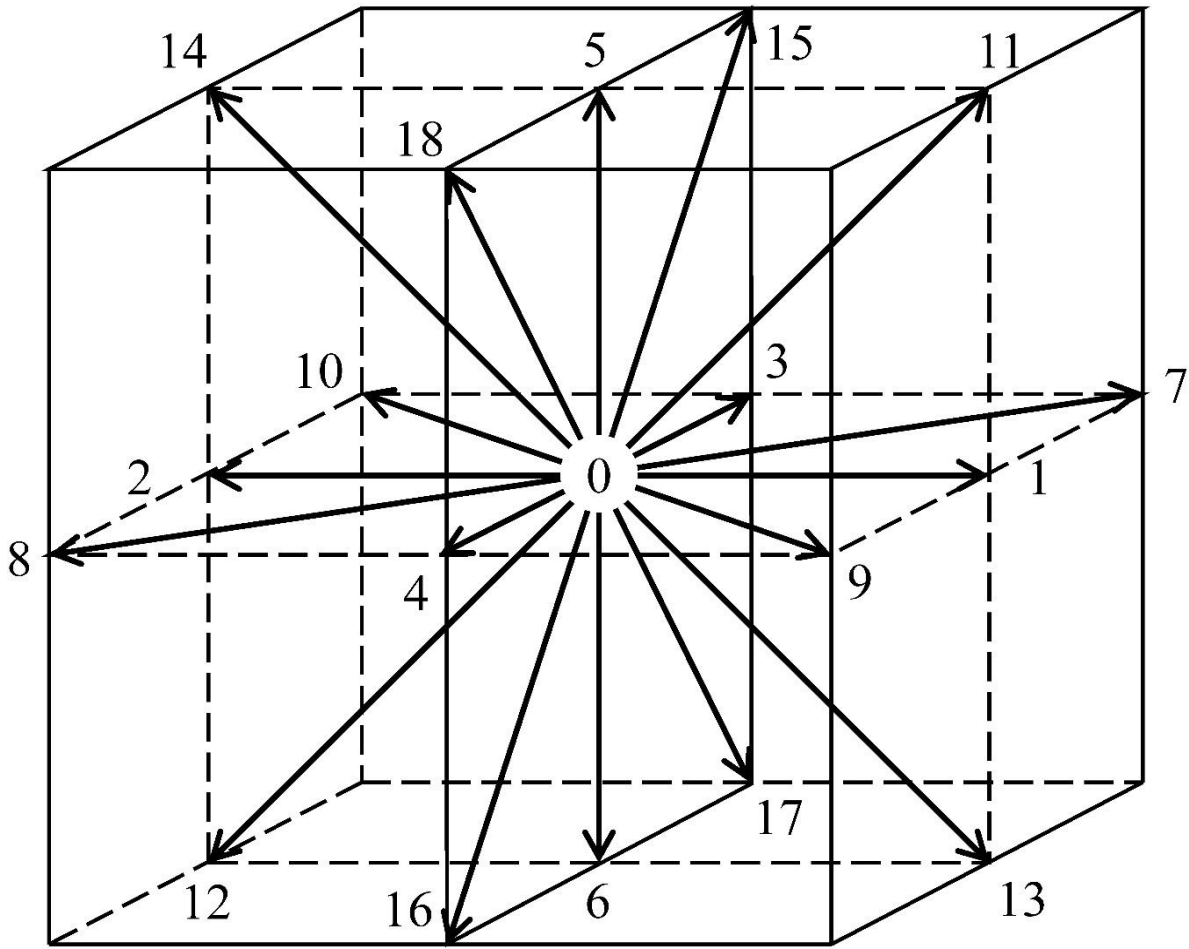

Fig.1 The D3Q19 lattice speed model.

The update of the density distribution functions $f_i(\mathbf{x},t)$ follows the lattice Boltzmann equation (LBE) [9],

$$f_i(\mathbf{x}+\mathbf{e}_i\Delta t, t+\Delta t) = f_i(\mathbf{x},t) + \Omega_i[f_i(\mathbf{x},t)], \tag{1}$$

where the vector **x** denotes the lattice position, $\Delta t$ is the time step, and $\Omega_i[f_i(\mathbf{x},t)]$ is the collision operator that controls the relaxation rate of the density distribution functions $f_i(\mathbf{x},t)$. The single-relaxation-time model of the collision operator is adopted [8,10,47-48], which is given as

$$\Omega_i = -\frac{\Delta t}{\tau}[f_i(\mathbf{x},t) - f_i^{eq}(\mathbf{x},t)] + F_i \Delta t, \qquad (2)$$

where $\tau$ is the dimensionless relaxation parameter and $F_i$ represents the body force acting on the fluid. $f_i^{eq}(\mathbf{x},t)$ is the equilibrium distribution function,

$$f_i^{eq}(\mathbf{x},t) = \rho \omega_i [1 + \frac{\mathbf{e}_i \cdot \mathbf{u}}{c_s^2} + \frac{(\mathbf{e}_i \cdot \mathbf{u})^2}{2c_s^4} - \frac{u^2}{2c_s^2}], \qquad (3)$$

where $\omega_i$ is the weight coefficient based on the lattice speed model. For the D3Q19 model, $\omega_0 = 1/3$, $\omega_{1\sim6} = 1/18$, $\omega_{7\sim18} = 1/36$. $c_s = c/\sqrt{3}$ is the lattice sound speed, where $c = |\Delta x/\Delta t|$ is the lattice speed. The body force $F_i$ in Eq. (2) is given in the following form, in order to correctly recover the Navier-Stokes equation with a body force term [49],

$$F_i = (1 - \frac{1}{2\tau})\omega_i [\frac{\mathbf{e}_i - \mathbf{u}}{c_s^2} + \frac{(\mathbf{e}_i \cdot \mathbf{u})}{c_s^4}\mathbf{e}_i] \cdot \mathbf{F}, \qquad (4)$$

where **F** is the macroscopic body force.

Then the macroscopic fluid properties, including fluid density, velocity, pressure and kinematic viscosity, can be determined by the microscopic density function and the discretisation parameters as

$$\begin{aligned}
\rho &= \sum_i f_i, \\
\rho \mathbf{u} &= \sum_i f_i \mathbf{e}_i + \frac{\Delta t}{2} \mathbf{F}, \\
p &= c_s^2 \rho, \\
\nu &= \frac{1}{3}(\tau - \frac{1}{2}).
\end{aligned} \qquad (5)$$

## 2.2 Boundary conditions

In LBM, the boundary conditions are implemented using the density distribution functions. In the current work, the 'no-slip' wall boundary conditions, the periodic boundary conditions and the moving wall boundary conditions are considered. The 'no-slip' boundary condition at the interface between the fluid and the stationary solid wall is implemented by the so-called bounce-back rule [14], which simply reverses the incoming density distribution functions from the fluid node back to the directions from which they come at all wall boundary nodes. The bounce-back rules can be defined as

$$f_{-i}(\mathbf{x}, t + \Delta t) = f_i^+(\mathbf{x}, t), \qquad (6)$$

where the subscript "-$i$" denotes the opposite direction to $i$ and $f_i^+$ represents the post-collision density distribution function. Note that the second-order accuracy can be achieved when the collision and streaming processes are also carried out at the solid boundary nodes [50].

Periodic boundary conditions are implemented in LBM in such a way that the density distribution functions exiting the domain at one end are duplicated to a virtual node at the other end. Then a normal streaming process takes place between the virtual node and the corresponding nodes at the other end. It should be noted that the solid particles immersed in the fluid are treated as moving wall boundaries. The interactions between the moving particles and the fluid play important roles in LBM-DEM coupling, which will be discussed separately in the next section.

## 2.3 Immersed boundary method with the partially solid scheme

In order to impose the correct 'no-slip' boundary conditions, the solid particle's boundaries are first mapped onto the lattice nodes. Figure 2(a) illustrates the lattice discretisation of a circular particle, where the nodes are further classified into three categories: (1) fluid boundary node; (2) solid boundary node; and (3) interior solid node. Obviously, the zigzag fashion of the particle surface is neither accurate nor smooth unless a sufficiently small lattice spacing is used.

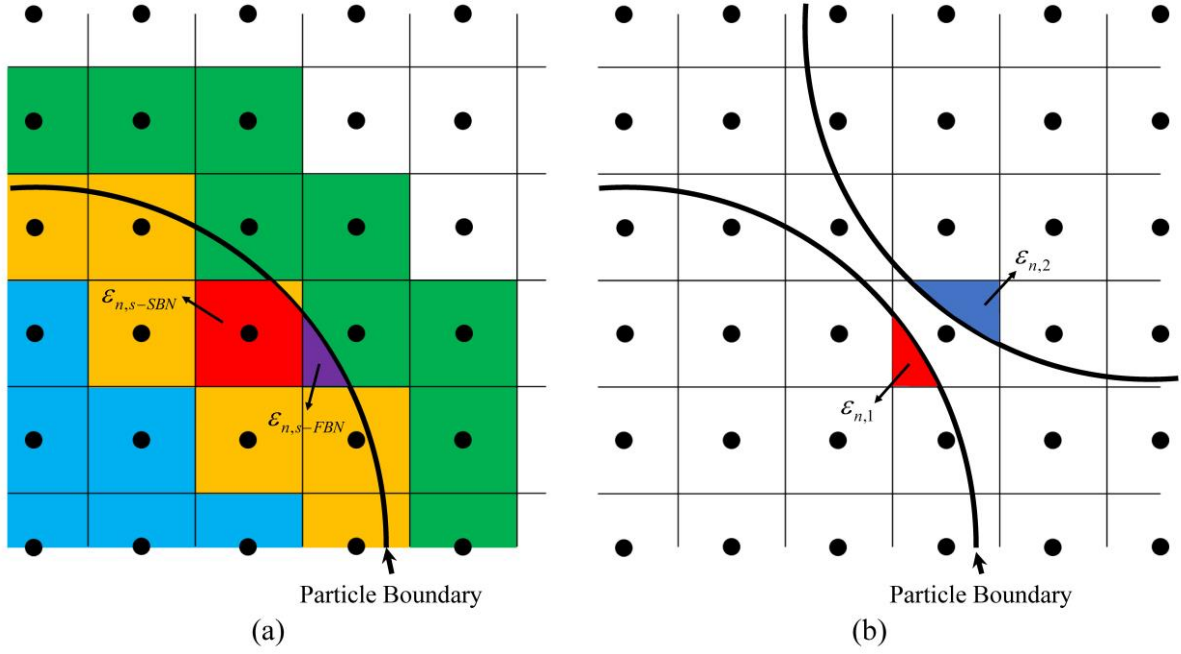

Fig. 2 (a) Lattice representation of a circular solid particle showing solid boundary nodes (orange), fluid boundary nodes (green) and internal solid nodes (blue). The red and purple parts represent the solid coverage ratio at a solid boundary node (SBN) and a fluid boundary node (FBN), respectively. (b) Schematic shows that two particles get in close contact and intersect with the same cell.

Based on the lattice representation, an immersed boundary technique based on the local solid fraction for LBM-DEM coupling was proposed by Noble and Torczynski [27], aiming to overcome the momentum discontinuity of BB-based techniques and to provide adequate representation of non-conforming boundaries at lower grid resolutions. It was also important to retain two critical advantages of the LBM, namely the locality of the collision operator and the simple linear streaming operator, and thus to facilitate solutions involving large numbers of irregular shaped moving boundaries. In this method, the lattice Boltzmann equation is modified to include a term that is dependent on the solid coverage ratio of the cell (see Fig. 2(a)), thus improving the boundary representation and smoothing the hydrodynamic forces calculated at the boundary nodes of a particle as it moves relative to the grid.

In the IBM, the modified lattice Boltzmann equation can be written to include the body force term as

$$f_i(\mathbf{x}+\mathbf{e}_i\Delta t, t+\Delta t) = f_i(\mathbf{x},t) - (1-B_n)[\frac{\Delta t}{\tau}(f_i(\mathbf{x},t) - f_i^{eq}(\mathbf{x},t))] + B_n\Omega_i^s + (1-B_n)F_i\Delta t, \quad (7)$$

where $\Omega_i^s$ is the additional collision term, and $B_n$ is a weighting function based on the total solid fraction in each cell, $\varepsilon_n$. Note that in the original method, both the solid fraction and the weighting function are summations of all coverage contributions from solid particles that intersect the same cell so that

$$\begin{aligned} \varepsilon_n &= \sum_s \varepsilon_{n,s}, \\ B_n &= \sum_s B_{n,s}, \end{aligned} \quad (8)$$

where $\varepsilon_{n,s}$ and $B_{n,s}$ are the contributions from each solid particle in the same cell. The additional collision term modifies the momenta of mapped particle nodes and accounts for fluid interaction with any solid particles in the cell. The weighting function is suggested to have the following form [27]

$$B_{n,s}(\varepsilon_{n,s}, \tau) = \frac{\varepsilon_{n,s}(\tau/\Delta t - 0.5)}{(1-\varepsilon_{n,s}) + (\tau/\Delta t - 0.5)}. \quad (9)$$

As the value of the solid fraction $\varepsilon_{n,s}$ varies from 0 (a completely fluid cell) to 1 (a completely solid cell), $B_{n,s}$ also varies from 0 to 1. Eq. (7) recovers to the original collision equation for pure fluid when $B_{n,s}=0$, and the new collision operator $\Omega_i^s$ plus the distribution from the previous time step when $B_{n,s}=1$. The new collision operator is then given as

$$\Omega_i^s = f_{-i}(\mathbf{x},t) - f_{-i}^{eq}(\rho,\mathbf{u}) + f_i^{eq}(\rho,\mathbf{u}_s) - f_i(\mathbf{x},t), \qquad (10)$$

where $\mathbf{u}_s$ is the velocity of the solid particle. The collision operator $\Omega_i^s$ is derived based on the bounce-back of the non-equilbrium part of the distribution functions. When $B_n=1$, the last term in Eq. (10) cancels the first term in Eq. (7), and the right-hand side of Eq. (7) only consists of the first three terms in Eq. (10). The term $f_{-i}(\mathbf{x},t) - f_{-i}^{eq}(\rho,\mathbf{u})$ bounces back the non-equilibrium part of the density function, and the term $f_i^{eq}(\rho,\mathbf{u}_s)$ is the equilibrium distribution function of the solid velocity. Note the straightforward implementation of the method: a single term ($\Omega_i^s$) is added to the LBE and two coefficients (1-$B_n$) in the equation are modified. Only quantities already available on the mesh or easily derived are used. No additional data storage or organisation is needed, which is a crucial issue in most moving boundary formulations. Calculations of the standard LBE for lattice nodes that are partially or completely covered by solids are replaced by the modified Eq. (7). The total hydrodynamic force and torque acting on the solid particle is determined by summing the change of momenta due to the additional collision operator over all lattice directions at each node and then over all fluid boundaries, solid boundaries and internal solid nodes, which are expressed as

$$\mathbf{F}_f = -\sum_n B_n(\sum_i \Omega_i^s \mathbf{e}_i),$$
$$\mathbf{T}_f = -\sum_n [(\mathbf{x}_n - \mathbf{X}_p) \times B_n(\sum_i \Omega_i^s \mathbf{e}_i)].$$
(11)

Here, $\mathbf{x}_n$-$\mathbf{X}_p$ is the vector from the center of rotation to the coupled node and the minus sign represents the direction of the force and torque according to the Newton's third law.

However, the original method becomes inaccurate when more than one particle intersects with the same cell, as shown in Fig. 2(b). As $B_{n,s}$ is non-linearly dependent on $\varepsilon_{n,s}$ (see Eq. (9)), the sum of $B_{n,s}$ does not equal to the value that is calculated using the sum of solid fraction $\varepsilon_{n,s}$, i.e. $B_n = \sum B_{n,s}(\varepsilon_{n,s}, \tau) \neq B_{n,s}(\sum \varepsilon_{n,s}, \tau)$, which leads to an incorrect weighting function in the solid-fluid coupling term. For example, consider that two particles intersect with the same cell with solid fractions of $\varepsilon_{n,1}$=0.4 and $\varepsilon_{n,2}$=0.6, respectively. The total solid fraction in this cell is $\varepsilon_n = \varepsilon_{n,1} + \varepsilon_{n,2} = 1$, indicating a completely solid cell. Then the corresponding total weighting function $B_n$ should be physically equal to one. Nevertheless, according to Eq. (9), the total weighting function is only $B_n$=0.18, 0.39, 0.52, given that $\tau/\Delta t$=0.6, 0.8, 1.0, respectively. Figure 3(a) shows the variations of the weighting function $B_{n,s}$ with the solid fraction of a single particle $\varepsilon_{n,s}$. It is clear that that $B_{n,s}$ increases non-linearly with $\varepsilon_{n,s}$. Only with a relatively large relaxation parameter, the relationship between $B_{n,s}$ and $\varepsilon_{n,s}$ approaches linear. Figure 3(b) further presents a comparison between the two different ways of calculating the total weighting function using Eq. (8) and Eq. (9), for the case that two particles intersect with the same cell, i.e. $B_n = B_{n,1}(\varepsilon_{n,1}, \tau) + B_{n,2}(\varepsilon_n - \varepsilon_{n,1}, \tau)$ and $B_n = B_{n,s}(\varepsilon_n, \tau)$. Three different total solid fractions are used for the calculation, $\varepsilon_n$=0.6, 0.8, 1.0, and the relaxation parameter is fixed at $\tau$=0.8. It can be seen that the original IBM method always underestimates the total weighting function. This distinction is substantially caused by the linear inconsistency in the definition of $B_n$ and $B_{n,s}$. Moreover, in the force calculation, the contribution to the force on each particle should also be weighted based on its solid fraction, which is actually not fully considered in the original IBM method. Therefore, an improved IBM method is proposed in the current study to address this problem.

The modified IBM-LBE is rewritten as

$$f_i(\mathbf{x}+\mathbf{e}_i\Delta t, t+\Delta t) = f_i(\mathbf{x},t) - (1-B_n)[\frac{\Delta t}{\tau}(f_i(\mathbf{x},t)-f_i^{eq}(\mathbf{x},t))] + \sum_s B_{n,s}\Omega_i^s + (1-B_n)F_i\Delta t, \quad (12)$$

where the term $B_n\Omega_i^s$ is replaced with the sum of the contribution $\sum_s B_{n,s}\Omega_i^s$ from each particle that intersects with the cell. $B_n$ keeps the same form as that in the original method, but $B_{n,s}$ is re-defined as

$$B_{n,s}(\varepsilon_{n,s},\tau) = \frac{\varepsilon_{n,s}(\tau/\Delta t - 0.5)}{(1-\varepsilon_n)+(\tau/\Delta t - 0.5)}, \quad (13)$$

where $\varepsilon_n$ is still the total solid fraction in the same cell. In this case, when there is only one particle that intersects with a cell, Eqs. (12) and (13) become identical to the original method. When there is more than one particle intersecting with a cell, their contributions to the total weighting function $B_n$ can be linearly summed up, which ensures that $\sum B_{n,s}(\varepsilon_{n,s},\tau) = B_{n,s}(\sum \varepsilon_{n,s},\tau)$. The additional collision term $\Omega_i^s$ also keeps the same form as in Eq. (10) but is updated with the velocity of corresponding solid particle. Similarly, the equations used to compute the force and torque are re-written as

$$\begin{aligned}\mathbf{F}_f &= -\sum_n B_{n,s}(\sum_i \Omega_i^s \mathbf{e}_i), \\ \mathbf{T}_f &= -\sum_n [(\mathbf{x}_n - \mathbf{X}_p) \times B_{n,s}(\sum_i \Omega_i^s \mathbf{e}_i)],\end{aligned} \qquad (14)$$

where the total weighting function $B_n$ is replaced by the weighting function $B_{n,s}$ of the corresponding particle in the cell. These modifications can accurately describe the forces in the cell with multiple particles, which is overestimated by the original model as a larger weighting function is applied. Even though this circumstance does not occur often in the dilute particle-liquid flows or when the particle-lattice size ratio is relatively large. However, it becomes quite important for the dense particle-liquid flows, where numerous close contacts between particles are present. The overestimation of the forces in the cells with multiple particles will be accumulated and can result in irrational numerical results.

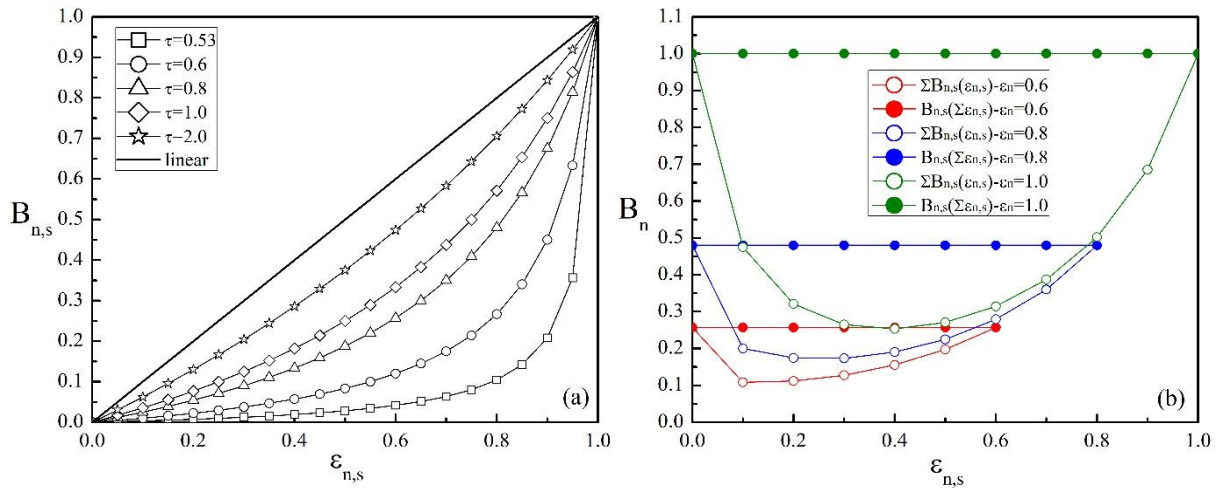

Fig. 3 (a) Weighting function $B_{n,s}(\varepsilon_{n,s}, \tau)$ as a function of the solid fraction $\varepsilon_{n,s}$ for different relaxation parameters $\tau$. (b) The total weighting function $B_n$ as a function of the solid fraction of a particle for the case that two particles intersect with the same cell. Comparison is made between the modified and the original weighting function $B_{n,s}$ for different total solid fractions $\varepsilon_n$. The relaxation parameter is fixed at $\tau=0.8$.

## 2.4 An accurate estimation of solid fraction in the lattice cell

In the immersed boundary method described above, it is clear that accurate calculation of the solid fraction in each lattice cell is crucial. In this section, a fast linear approximation method is introduced, which was recently proposed by Jones and Williams [51], to estimate the solid fraction of a single particle that intersects with a lattice cell. Figure 4 shows the schematic of the linear approximation in a simplified 2D view for better illustration. In this approach, the solid fraction is simply approximated by a linear function,

$$\varepsilon_s = -D + f(r), \qquad (15)$$

where $D$ is the distance from the cell centre to the particle surface. $f(r)$ is a function of the normalised particle radius. Eq. (15) is derived from the analytical solution to the intersection volume calculation for a specific cell orientation with respect to the particle surface, based upon the following assumptions:

(*i*) The cell and particle centres lie along a single axis, i.e. they have a *z*-component of zero.

(*ii*) A pair of opposing faces in the cell is parallel to the plane tangential to the sphere surface.

(*iii*) The sphere surface intersects only the remaining cell faces, and not those parallel to the tangent plane.

As illustrated in Fig. 4, by setting the particle centre as the origin, the solid fraction equals to the intersection volume $V_i$, i.e.

$$\varepsilon_s = V_i = V_a - V_b, \tag{16}$$

where $V_b$ is the volume of the cuboid between the lower cell face and the origin, $V_b = y_c - 0.5$. $y_c$ is the y-coordinate of the cell centre. $V_a$ is calculated with the integral,

$$\begin{aligned}V_a &= \int_{-0.5}^{0.5}\int_{-0.5}^{0.5}(r^2 - x^2 - y^2)dxdy \\ &= (\frac{1}{12} - r^2)\tan^{-1}(\frac{0.5\sqrt{r^2-0.5}}{0.5-r^2}) + \frac{1}{3}\sqrt{r^2-0.5} \\ &+ (r^2 - \frac{1}{12})\tan^{-1}(\frac{0.5}{\sqrt{r^2-0.5}}) - \frac{4}{3}r^3\tan^{-1}(\frac{0.25}{r\sqrt{r^2-0.5}}),\end{aligned} \tag{17}$$

which is a constant value for a fixed particle radius. Note that $y_c = r + D$, then by substituting $V_a$ and $V_b$ into Eq. (16), we have

$$f(r) = V_a - r + 0.5. \tag{18}$$

This approach is also valid in 2D, where the integral to compute $V_a$ becomes

$$V_a = \int_{-0.5}^{0.5} (r^2 - x^2)dx = 0.5\sqrt{r^2 - 0.25} + r^2 \tan^{-1}(\frac{0.5}{\sqrt{r^2 - 0.25}}). \tag{19}$$

This linear approximation is proved very accurate when applied in the general case and significantly saves the computational cost [51]. However, according to the assumptions, some special cases still need to be handled. For instance, if the cell only intersects with the particle at a corner, Eq. (15) could give negative values due to the dissatisfaction of the assumptions. Therefore, in the actual implementation, we still use Eq. (15) to calculate the solid fraction in any case but force the solid fraction to be in the range of [0,1]. That is, $\varepsilon_s$=0 if Eq. (15) gives a negative value; $\varepsilon_s$=1 if Eq. (15) gives a value above one. Furthermore, it is obvious that the solid fractions of the cells completely inside and outside the particle are one and zero, respectively. Therefore, a further speed up of the algorithm can be achieved by a simple search with the distance from the cell centre to the particle centre. Specifically, in this current work, only the lattice cells with the centre-to-centre distance in the range [$r-h$, $r+h$] are computed with the linear approximation method, which are contained in a spherical shell with the thickness of 2$h$ lattice units, with $h$ being a tunable variable. Usually, $h$ can be set to one as the particle diameter is at least 5 lattice units in the LBM-DEM. Table 1 summarises the algorithm of the calculation of the solid fraction.

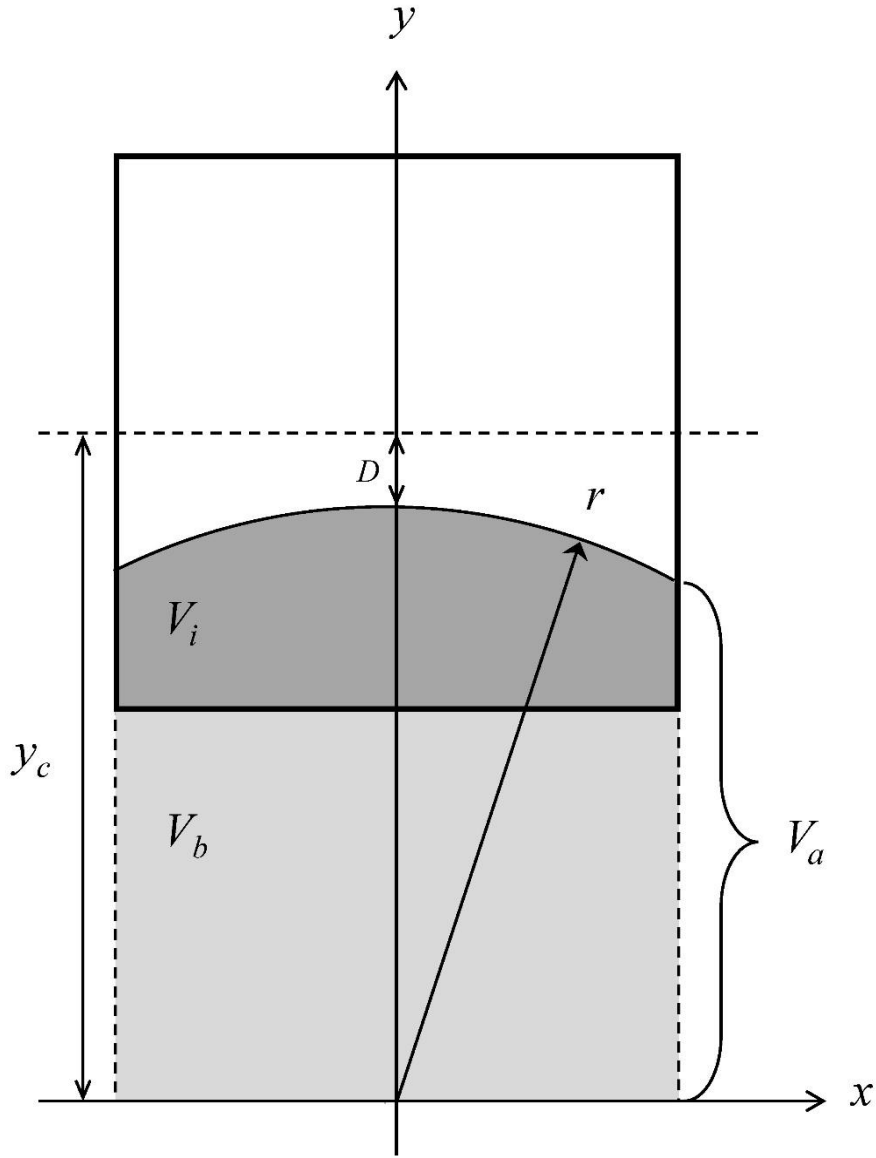

Fig. 4 Schematic of linear approximation in 2D.

Table 1 Algorithm of the calculation of solid fraction in a lattice cell

---

DO n=1, Num_of_Particle (loop over all the solid particles)

Step 1      Find all the lattices that fully contain the current solid particle.

         DO m=1, Num_of_Lattice (loop over all the lattices found in Step 1)

Step 2      Calculate the distance between the lattice cell and center of particle.

Step 3    IF the distance $\geqslant$ particle radius + h

   Solid fraction = 1.0

   ELSE IF the distance $\leqslant$ particle radius − h

   Solid fraction = 0.0

   ELSE IF the distance is in between

   Use Eqs. (15), (16) and (18) to calculate the solid fraction.

   END IF

Step 4    Modify the unreasonable values of solid fraction calculated by the linear approximation equations.

   IF solid fraction $\leqslant$ 0.0, THEN solid fraction = 0.0

   IF solid fraction $\geqslant$ 1.0, THEN solid fraction = 1.0

Step 5    Calculate the total solid fraction in the current cell and modify the unreasonable values of total solid fraction.

   IF total solid fraction $\geqslant$ 1.0, THEN total solid fraction = 1.0

Step 6    Record the particle ID that intersects with the cell.

   END DO (move on to next lattice cell)

END DO (move on to next solid particle)

To assess the accuracy and efficiency of the linear approximation method, we calculate the solid fraction of a single sphere with increasing particle-to-lattice size resolutions and compare the results with other methods, including the Monte Carlo method and the polyhedron approximation method [51]. The Monte Carlo method is a statistical sampling technique, which randomly places a number of sampling points with uniform distribution in the lattice cell. The distance between each sample point and the sphere centre is calculated and compared with the sphere radius. If the distance is smaller than the radius, the corresponding sample point is counted as an inside point. Then the intersecting solid fraction is estimated as the ratio of the number of the inside points to the total number of sample points. Obviously, the more sample points are placed, the more accurate the estimation is, but the longer time the computation takes. The polyhedron approximation is a simplified analytical computational method, in which the intersecting solid fraction is approximated as a combination of polyhedra in 3D or a polygon in 2D. The key steps in this method is to identify the polyhedron that is defined by the intersecting points at the edges of the lattice cell. Generally, the larger the sphere-to-lattice size ratio is, the more accurate the approximation is, because the volume of the unaccounted spherical cap in the approximation becomes smaller.

For the test setup, we place a sphere in the centre of a cuboid box with the size of $(d+5)^3$, where $d$ is the sphere diameter in lattice unit and increases from 10 to 100, corresponding to the total number of lattice cells from $10^3$ to $10^6$. The computation time is estimated on a PC with the configuration of Intel Core i7-6700, 4 cores (8 threads), 3.4 GHz and 16 GB RAM, and without any parallel computing. Upon the calculation of the solid fraction in every lattice cell, the relative error in the total volume of the sphere can be estimated as

$$Err = \frac{|V_{p,c} - V_{p,a}|}{V_{p,a}} \times 100\% . \tag{20}$$

where $V_{p,c}$ represents the computed total volume and $V_{p,a} = \pi d^3/6$ is the sphere volume. Figure

5 shows the comparison of the computational performance for different methods. It is clear that the computational speed of the Monte Carlo (MC) method is of the order $O(n)$ and $O(d^3)$, where $n$ denotes the total number of the sample points, while the relative error decreases less than 3 orders of magnitude as the resolution increases by an order of 1. Besides, the accuracy of MC seems to be saturated when $n \geq 1,000$. For the other two methods, the computational speeds of the lineal approximation and the polyhedron approximation are almost the same, both of which are much faster than the MC. The computational time of these two methods is around the order of $O(d^2)$, increasing with the number of grid cells on the particle surface. However, the relative error of the linear approximation is generally an order of magnitude smaller than that of the polyhedron approximation. Although the relative error is still one order higher than the MC, the overall performance of the linear approximation is very satisfactory, considering the fast computational speed and the acceptable relative error. A more detailed comparison of the computational performance of the linear approximation with other available methods, including the full analytical solution method, the cell decomposition method, and the edge-intersection averaging method, can be found in [51].

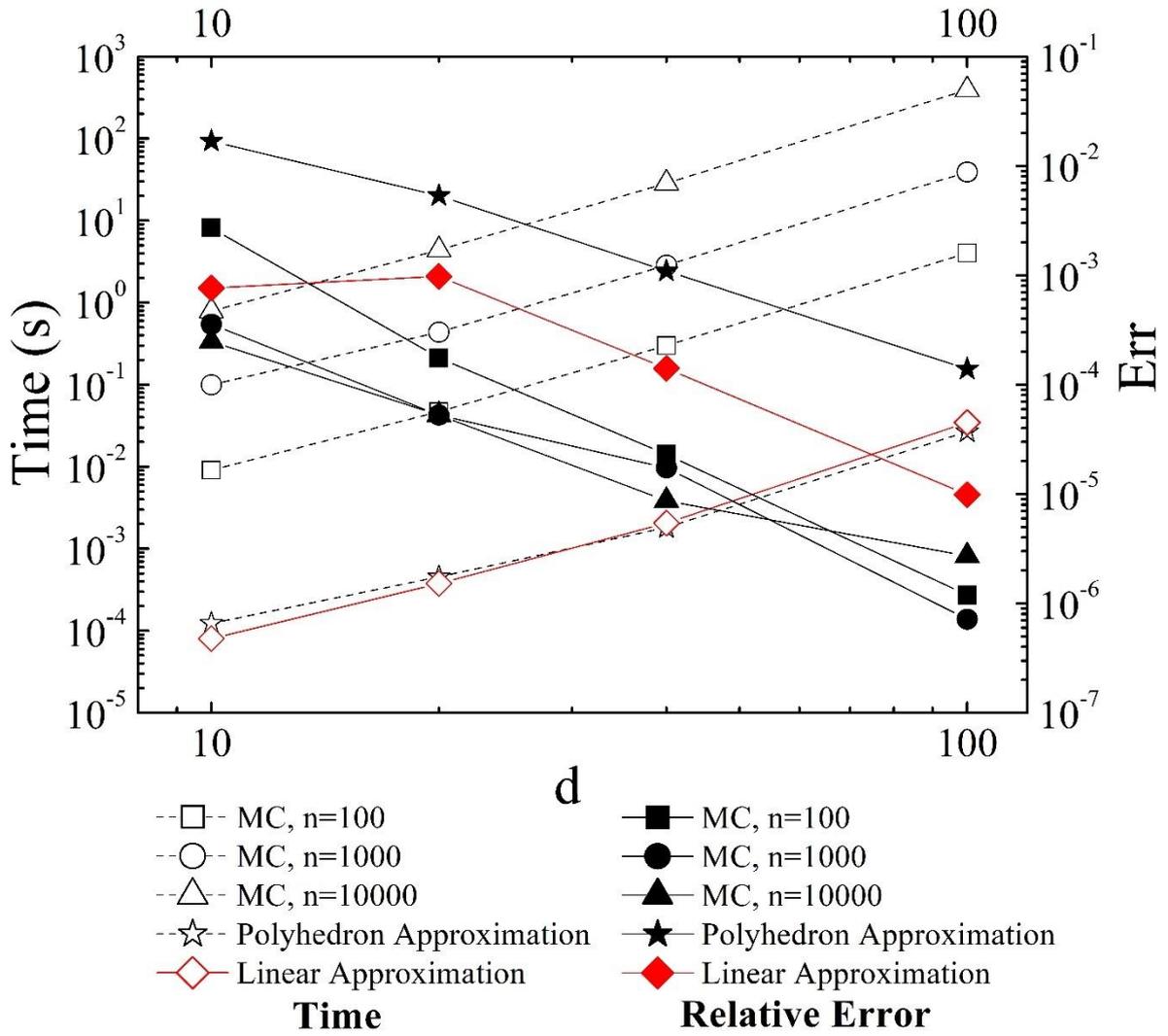

Fig. 5 Comparisons of the computational performance of different methods for calculating the solid fraction.

## 2.5 The discrete element method for adhesive particles

In DEM, the particle's motion is described by the Newton's second law [40-41], i.e.

$$m\frac{d\mathbf{U}_s}{dt} = \mathbf{F}_f + \mathbf{F}_c + \mathbf{G},$$
$$I\frac{d\mathbf{\Omega}_s}{dt} = \mathbf{M}_f + \mathbf{M}_c, \tag{21}$$

where $\mathbf{U}_s$ and $\mathbf{\Omega}_s$ are the transitional velocity and the rotational velocity of an individual particle, respectively. $m$ is the particle mass and $I$ is the moment of inertia. $\mathbf{G}$ is the gravitational force. $\mathbf{F}_f$ and $\mathbf{M}_f$ denote the fluid force and torque acting on each individual particle, respectively, and $\mathbf{F}_c$ and $\mathbf{M}_c$ are the force and torque resulted from the interparticle contact, respectively. The contact force and torque can be decomposed as

$$\mathbf{F}_c = F_n\mathbf{n} + F_s\mathbf{t}_s,$$
$$\mathbf{M}_c = r_p F_s(\mathbf{n}\times\mathbf{t}_s) + M_r(\mathbf{t}_r\times\mathbf{n}) + M_t\mathbf{n}, \tag{22}$$

where $F_n$ is the normal force including the elastic contact force and the damping force, $F_s$ is the tangential force due to the sliding friction, $M_r$ is the rolling resistance and $M_t$ is the twisting resistance. $r_p$ is the particle radius. $\mathbf{n}$, $\mathbf{t}_s$ and $\mathbf{t}_r$ are the normal, tangential and rolling direction unit vectors, respectively.

In a collision process between two adhesive microparticles, the particles undergo the jump-on and pull-off processes when they contact with and detach from each other, respectively, which greatly differ from the collision of granular particles [41]. As shown in Fig. 6, at the jump-on point (point A), the contact region area suddenly goes from zero to a finite value and the contact force suddenly goes from zero to a negative value, which leads to a first-contact energy loss. The necking behaviour of the material when the particles are pulled away from each other allows the adhesive force to act even when the normal overlap becomes negative, resulting in another an energy loss. The particles will finally detach from each other when a critical pull-off force is reached (point C). The energy dissipation due to the jump-on and pull-off behaviour of the adhesive contact is estimated as [52]

$$\Delta E_{ad} = \int_{-\delta_C}^{0} F_{ad} d\delta_N = 22.51 (\frac{\gamma^5 R^4}{E^2})^{1/3}. \tag{23}$$

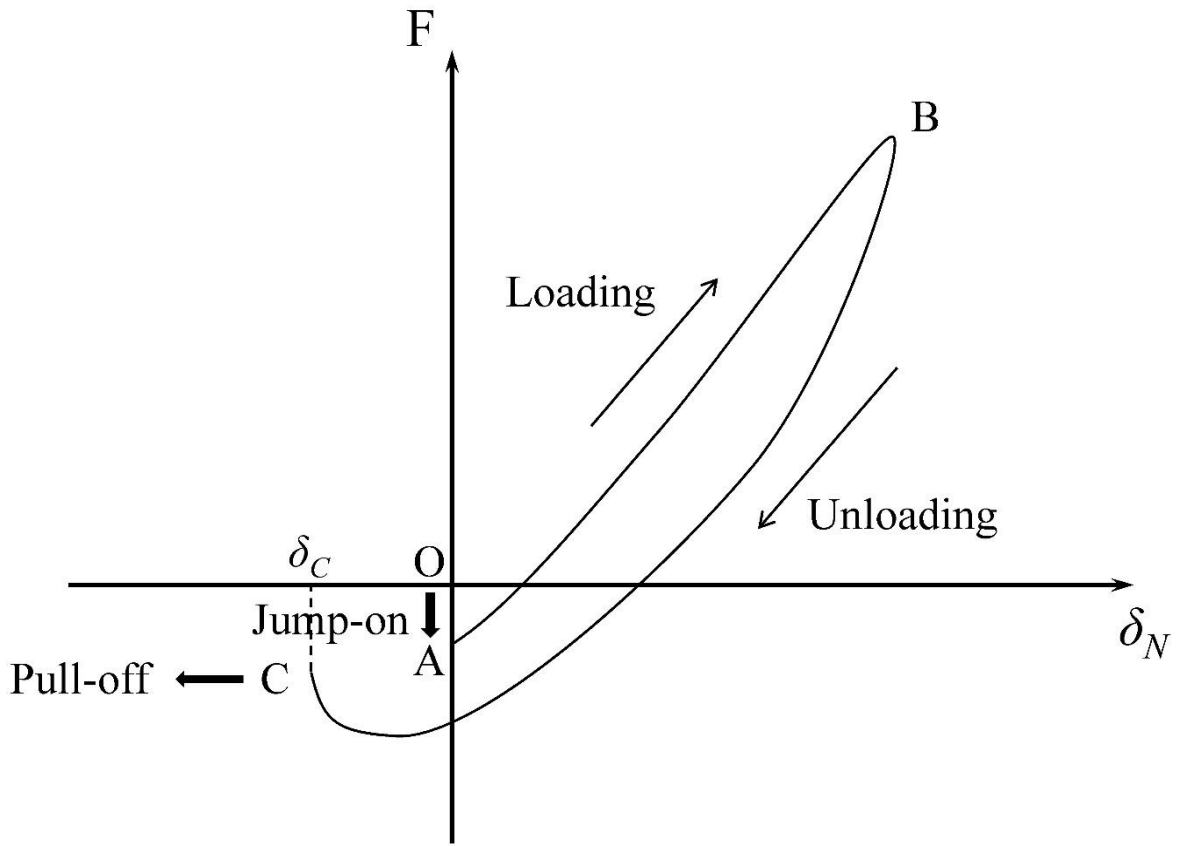

Fig. 6 Force-displacement curve during the normal loading and unloading of two adhesive spheres.

The other part of the energy loss arises from the solid-phase damping caused by the viscoelasticity of materials, which can be calculated as [40,53]

$$\Delta E_d = \int_{-\delta_C}^{\delta_{\max}} F_{nd} d\delta_N. \tag{24}$$

In Eqs. (23) and (24), $F_{ad}$ and $F_{nd}$ represent the adhesive force and damping force in the normal direction between two microparticles, respectively. $\delta_N$ is the normal overlap that equals $\delta_C$ at the critical pull-off point, and $\gamma$ is the surface energy. $R$ is the effective radius and $E$ is defined as the effective elastic modulus between two contacting particles,

$$\begin{aligned}\frac{1}{R} &= \frac{1}{r_{p,i}} + \frac{1}{r_{p,j}}, \\ \frac{1}{E} &\equiv \frac{1-\sigma_i^2}{E_{p,i}} + \frac{1-\sigma_j^2}{E_{p,j}},\end{aligned} \tag{25}$$

where $r_p$, $E_p$ and $\sigma$ denote the particle radius, elastic modulus and Poisson's ratio, respectively, and the subscripts $i$ and $j$ correspond to the two particle indexes. Note that we do not consider the plastic deformation in the present work, because it only becomes non-negligible when the impact velocity of particles is much larger than their critical sticking/rebound velocity [40,53]. Combing these two energy dissipation mechanisms, the instantaneous normal contact force $F_n$ is given by

$$F_n = F_{ad} + F_{nd} = 4F_C\left[\left(\frac{a}{a_0}\right)^3 - \left(\frac{a}{a_0}\right)^{3/2}\right] + \eta_N \mathbf{v}_R \cdot \mathbf{n}, \tag{26}$$

where the adhesively normal contact force is described by the JKR (Johnson-Kendall-Roberts)

model [54] and the damping force is assumed to be proportional to the rate of change of material deformation [40-41]. $F_C$ is the critical pull-off force derived from the JKR theory, $F_C=3\pi\gamma R$ [54], and $a$ is the radius of the contact area with $a_0$ being the equilibrium contact radius, which is given as

$$a_0 = \sqrt[3]{\frac{9\pi\gamma R^2}{E}}. \qquad (27)$$

$\eta_N$ is the normal dissipation coefficient, and $\mathbf{v}_R$ is the relative velocity at the contact point on particle surfaces. To minimize the computational time, $F_{ad}/F_C$ and $a/a_0$ as functions of $\delta_N/\delta_C$ are pre-computed at the beginning, and then we use a look-up table to determine $F_{ad}$ and $a(t)$ for the given value of $\delta_N$ at each time step.

Apart from the normal deformation, the interparticle sliding, twisting and rolling frictions are also considered and approximated with a linear spring-dashpot-slider model. They can be expressed as

$$\begin{aligned} F_s &= -\min[k_T(\int_{t_0}^{t}\mathbf{v}_R(\tau)\cdot\mathbf{t}_s\mathrm{d}\tau)+\eta_T\mathbf{v}_R\cdot\mathbf{t}_s, F_{s,crit}], \\ M_t &= -\min[\frac{k_T a^2}{2}(\int_{t_0}^{t}\Omega_t(\tau)\mathrm{d}\tau)+\frac{\eta_T a^2}{2}\Omega_t, M_{t,crit}], \\ M_r &= -\min[4F_C(\frac{a}{a_0})^{3/2}(\int_{t_0}^{t}\mathbf{v}_L(\tau)\mathrm{d}\tau), M_{r,crit}], \end{aligned} \qquad (28)$$

where $\mathbf{v}_R\cdot\mathbf{t}_s$, $\Omega_t$ and $\mathbf{v}_L$ stand for the relative sliding, twisting and rolling velocity, respectively. $k_T$ and $\eta_T$ are the tangential stiffness and dissipation coefficient, respectively. According to Eq.

(28), the sliding, twisting and rolling resistances first increase cumulatively with the increase of the corresponding displacements. Once reaching certain critical values, i.e. $F_{s,crit}$, $M_{t,crit}$ and $M_{r,crit}$, the resistances stay constant and the particles start to slide, spin or roll against each other. These critical values in the presence of adhesion are given in the following equations

$$\begin{aligned} F_{s,crit} &= \mu_f \left| F_{ne} + 2F_C \right|, \\ M_{t,crit} &= 3\pi a F_{s,crit}/16, \\ M_{r,crit} &= -4 F_C \left( a/a_0 \right)^{3/2} \theta_{crit} R, \end{aligned} \tag{29}$$

where $\mu_f$ is the friction coefficient and $\theta_{crit}$ is the critical angle for the relative rolling of two particles. The model parameters used in Eqs. (26) and (28) can be found in literature [53,55].

Regarding the time steps in the coupling of LBM and DEM, an efficient technique to handle any disparity in the time steps is necessary. The LBM time step is determined from the fluid kinematic viscosity of fluid, the grid resolution, and the relaxation parameter, while the DEM time step can be estimated by the Rayleigh wave speed of force transmission. Both time steps can vary in many orders of magnitude. Hence, we introduce a time step ratio $\lambda = \Delta t_{LBM}/\Delta t_{DEM}$. Before running the simulations, both LBM and DEM time steps are estimated based on the simulation setup in order to obtain the time step ratio. It should be noted that the time step ratio $\lambda$ is not necessarily greater than one. If $\Delta t_{LBM} < \Delta t_{DEM}$, the critical time step can be set as the LBM time step. Otherwise, a DEM sub-cycling is performed within one LBM time step. It is important to note that during DEM sub-cycling the hydrodynamic forces and torques are not updated.

# 3. Numerical verification

## 3.1 Duct flow in 3D

We firstly test our model with the fully developed laminar flow in a 3D duct. The flow channel is bounded by two pairs of parallel walls at the top, the bottom, the front and the back. Periodic boundary conditions are applied in the other two directions (inlet and outlet). The size of the channel is 10×51×51 and the fluid density, kinematic viscosity and relaxation parameter are 1,000 kg/m$^3$, 1×10$^{-4}$ m$^2$/s and 0.65, respectively. The flow is driven by a constant body force, which acts as the pressure gradient. By varying the body force between 3.2×10$^{-6}$ and 6.4×10$^{-5}$ (in lattice unit) various channel Reynolds numbers can be achieved. The velocity profile in a duct flow is analytically given as [56]

$$U(y,z) = \frac{G}{2\mu} y(l_h - y) - \frac{4Gl_h^2}{\mu\pi^3} \sum_{m=1}^{\infty} \frac{1}{(2m-1)^3} \frac{\sinh(\beta_m z) + \sinh(\beta_m (l_w - z))}{\sinh(\beta_m l_w)} \sin(\beta_m y), \quad (30)$$

where $\beta_m = \frac{2m-1}{l_h}\pi$, $l_h$ and $l_w$ stand for the height and width of the channel, respectively, and $G$ and $\mu$ represent the pressure gradient and fluid dynamic viscosity, respectively.

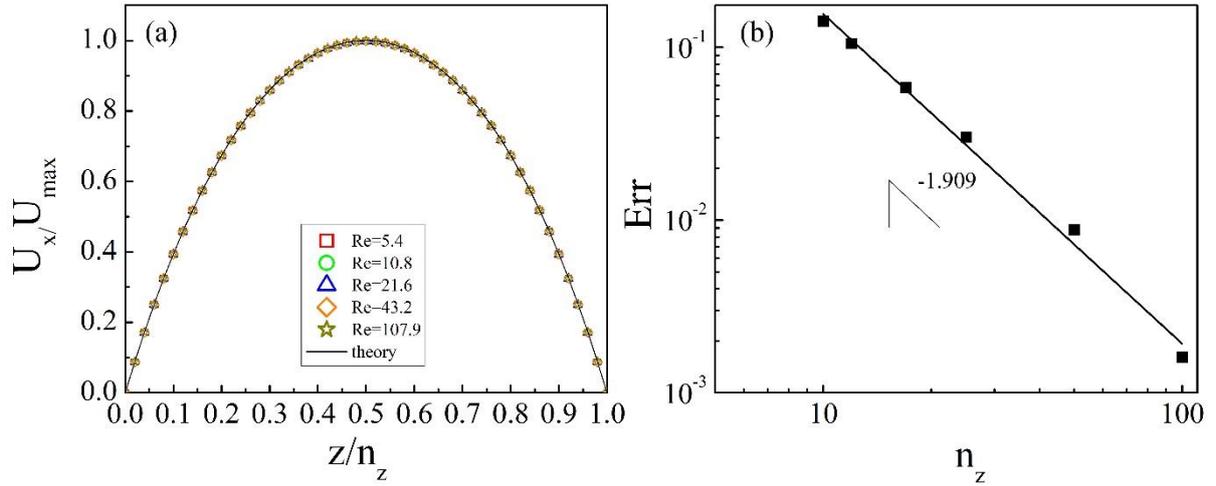

Fig. 7 (a) Normalized velocity profiles for duct flow with different Reynolds numbers at the position *x*=5, *y*=26. (b) The relative error of the flow velocity as a function of the number of lattice.

Figure 7(a) shows the velocity profiles for different channel Reynolds numbers, which are normalized with the corresponding maximum velocity along the centre line. It can be seen that the normalized velocity profiles for different channel Reynolds numbers all collapse onto a single curve, which agree perfectly with the theoretical prediction, within a maximum relative error less than 0.5%. Figure 7(b) shows the relative error as a function of the number of lattice, which is computed by means of an L-2 norm calculation

$$Err = \sum_n \sqrt{\frac{(U - U_{theory})^2}{U_{theory}^2}}, \qquad (31)$$

where the summation is taken over all the lattice nodes. As can be seen from Fig. 7(b), near second-order accuracy is attained. Therefore, it indicates that the present numerical approach is capable of accurately modelling the single-phase fluid flow.

## 3.2 Drag force on a fixed sphere

It was reported that the immersed moving boundary method could improve the boundary representation and smooth the hydrodynamic forces calculated at an obstacle's boundary nodes [34]. To test the accuracy in the force calculation, the flow past a fixed sphere is modelled and the drag force is calculated. The sphere is fixed in the centre of a cuboid box with size of $20d_p \times 10d_p \times 10d_p$, where $d_p$ is the diameter of the sphere. The box length is longer than 30 radii in order to eliminate the periodic effect in the flow direction [34]. The inlet boundary is constant flow with velocity $U_0$ and the outlet is set as constant pressure boundary. All the other boundaries are set as open boundary (zero gradient). The drag coefficient is calculated as

$$C_d = \frac{8F_d}{\rho_f U_0^2 \pi d_p^2}. \qquad (32)$$

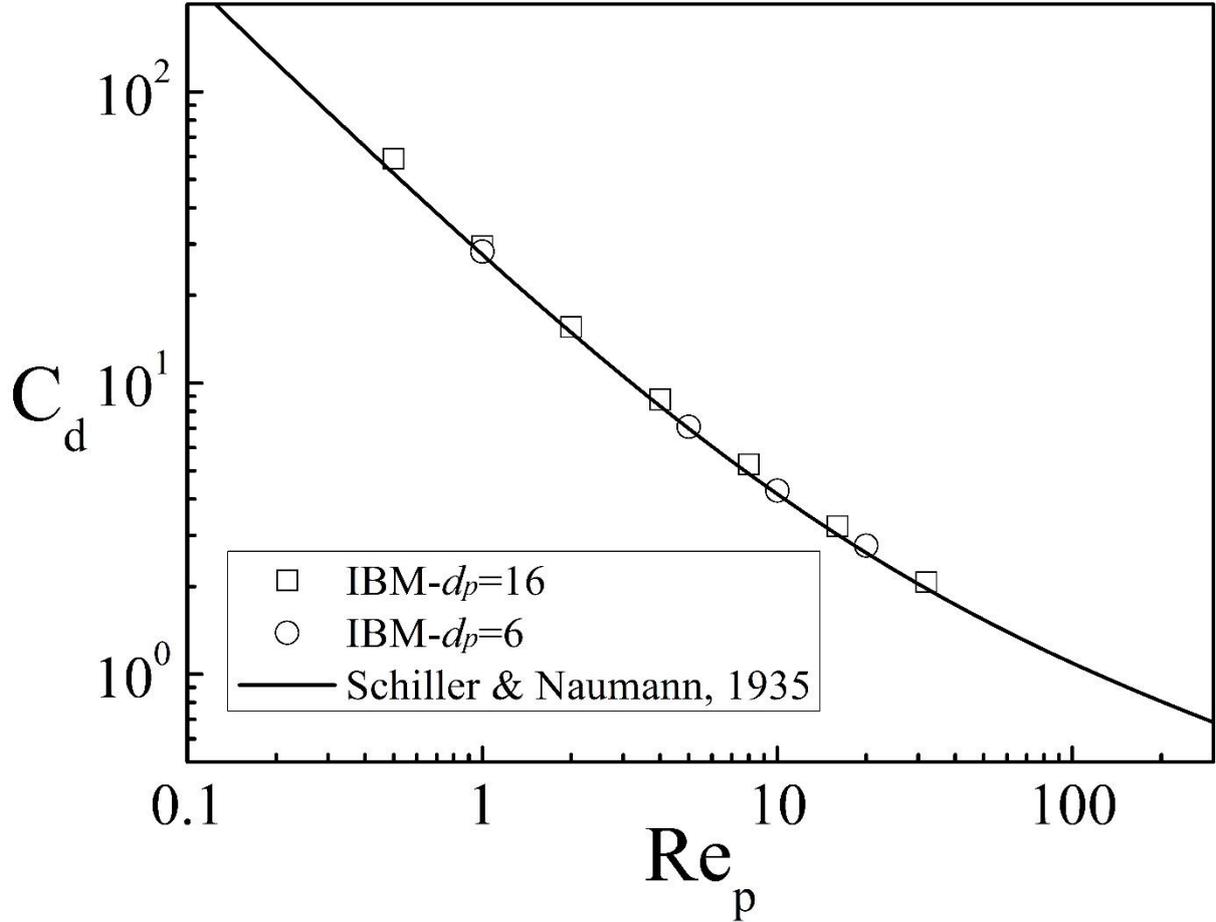

Fig. 8 Drag coefficient as a function of particle Reynolds number.

Figure 8 shows the drag coefficient $C_d$ as a function of the particle Reynolds number $Re_p$. The solid line refers to a widely accepted empirical law of the drag coefficient proposed by Schiller and Naumann [57],

$$C_d = \frac{24}{Re_p}(1+0.15\,Re_p^{0.687}), \tag{33}$$

which remains valid to within 5% of the experimental data for particle Reynolds number up to 800. It is clear that the IBM results agree well with the empirical equation, with a maximum relative error of 7.8%. Furthermore, it is noticed that the drag coefficient is quite accurate even with a relatively low size resolution $d_p$=6. As a consequence, it is concluded that the IBM method is accurate for the force calculation, which is able to provide acceptable accuracy with a reduced resolution.

## 3.3 Gravitational settling

In order to explore the computational accuracy of our numerical approach in a dynamic system, gravitational sedimentations in both 2D and 3D cases are analysed. For 2D simulations, the computational setup is identical to that in Wen et al.'s work [58]. A cylinder with diameter of 0.1 cm is initially released at 0.076 cm away from the centreline of a vertical channel with width of 0.4 cm. The fluid density and kinematic viscosity are 1,000 kg/m$^3$ and 1×10$^{-6}$ m$^2$/s. The mass density of the cylinder is 1,030 kg/m$^3$, which is slightly larger than the fluid's. Therefore, the cylinder rotates and translates under the gravitational and hydrodynamic forces and finally reaches a steady state settling along the centreline at a constant velocity. Two different lattice resolutions are selected as $d_p$=30 and $d_p$=15, where the larger one is the same as that in [58]. The dimensionless relaxation parameter is set as 0.6.

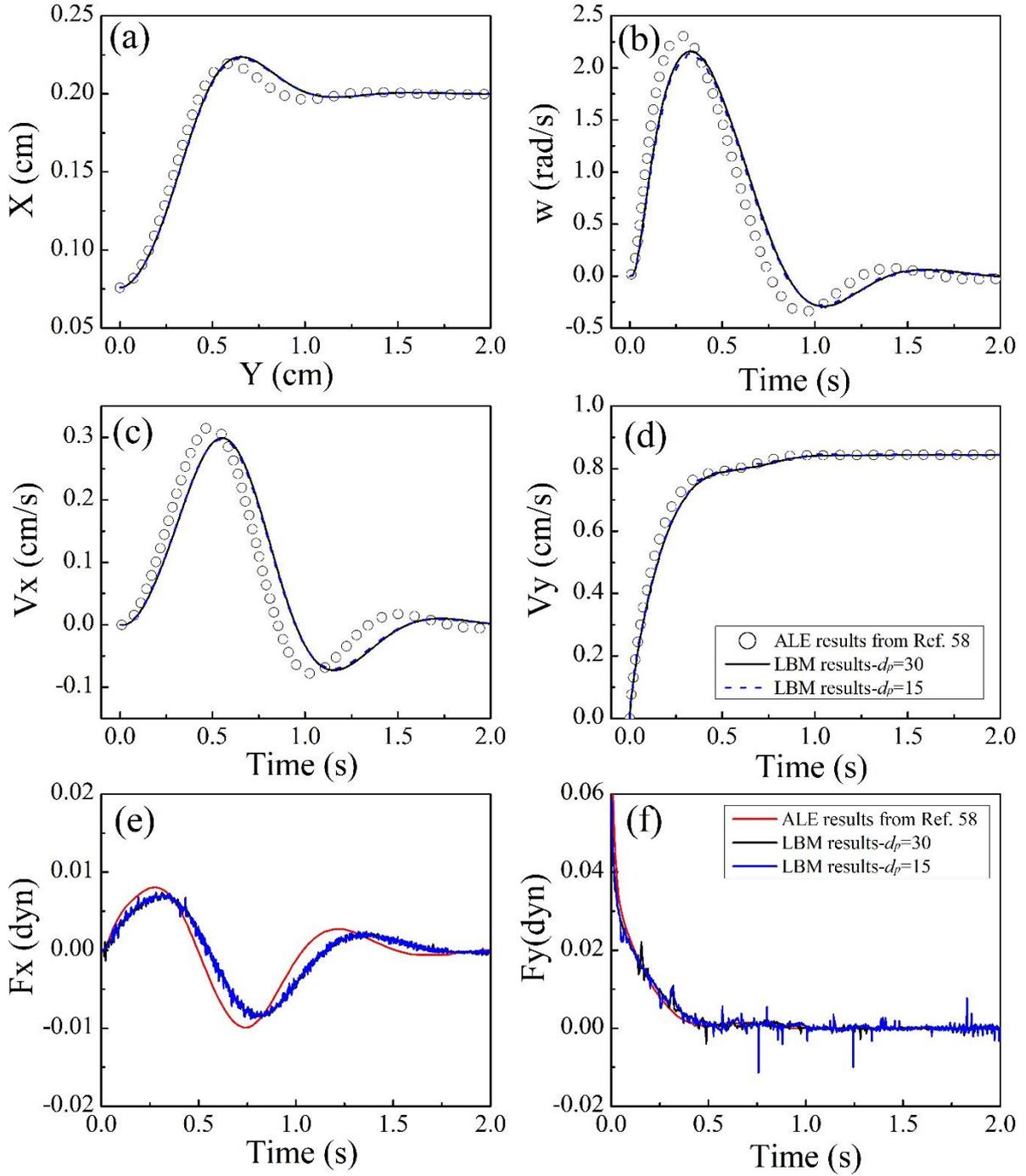

Fig. 9 Time-dependent (a) particle trajectory, (b) angular velocity, (c) horizontal velocity, (d) vertical velocity, (e) horizontal force and (f) vertical force.

The results are presented in Figure 9 together with the comparison with that obtained using the conventional equation and the arbitrary Lagrangian-Eulerian technique (ALE) [58]. Good agreement is found between our LBM results and the ALE results. The force profiles are quite smooth without any large fluctuations. Furthermore, it is also observed that the results of the relatively lower resolution $d_p=15$ are almost identical to those of $d_p=30$, which demonstrates the accuracy of the IBM at lower lattice resolution.

For 3D numerical simulations, the experiment of a single particle settling under gravity is reproduced numerically [59]. A particle is initially released from the height of 120 mm in a box with size of $100\times100\times160$ mm$^3$, which corresponds to $50\times50\times80$ in lattice units. The gravity is in the vertical direction and no-slip wall conditions are set in all the faces of the box. The relaxation parameter is set as 0.65. The fluid density and viscosity are identical to that used in the experiment (see Table I in [59]). The particle diameter is $d_p=15$ mm, which equals to 7.5 in lattice units, and the mass density is fixed at $\rho_p=1,120$ kg/m$^3$. Figure 10 shows the particle settling trajectory and velocity profiles as a function of time, which is converted to SI unit to compare with the experimental results. It is clear that, for both the settling trajectory and the velocity, the numerical simulations are in excellent agreement with the experimental results, which demonstrates the validity and accuracy of our numerical approach.

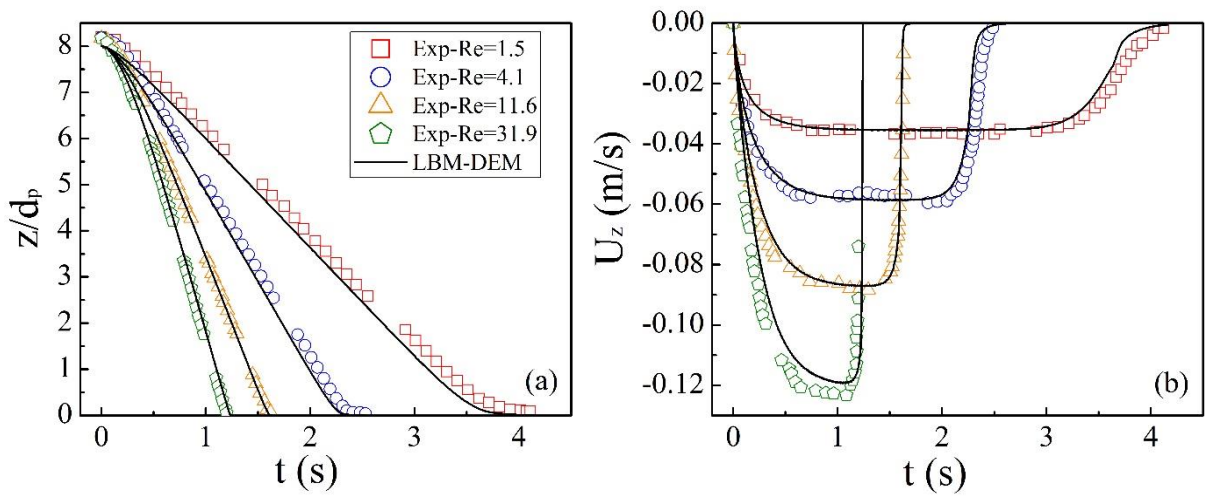

Fig. 10 (a) Particle trajectory and (b) velocity profiles as a function of time.

## 3.4 Flow through porous media

Simulations of the flow through porous media are also performed to further validate our improved IBM method when particles are in close contact. It is well accepted that the pressure drop in one-dimensional fluid flow through a fixed packed bed consisting of spherical particles can be accurately predicted by the Ergun equation [60],

$$\frac{\Delta P}{h} = \frac{150\mu U}{d_p^2}\frac{(1-\varepsilon)^2}{\varepsilon^3} + \frac{1.75\rho_f U^2}{d_p}\frac{(1-\varepsilon)}{\varepsilon^3}, \quad (34)$$

where $\Delta P$ is the pressure difference, $h$ is the height of the packed bed, $\mu$ is the fluid dynamical viscosity, $\rho_f$ is the fluid density, $U$ is the superficial fluid velocity, $d_p$ is the particle diameter, $\varepsilon$ is the porosity of the bed. The Ergun equation can be transformed to a dimensionless expression

$$\Delta P^* = \frac{150}{\text{Re}_p^*} + 1.75, \quad (35)$$

where $\Delta P^*$ is the dimensionless pressure loss and $Re_p^*$ is the modified Reynolds number, which are defined as

$$\Delta P^* = \frac{\Delta P}{h}\frac{d_p}{\rho_f U^2}\frac{\varepsilon^3}{(1-\varepsilon)^2},$$
$$\text{Re}_p^* = \frac{\rho_f d_p U}{\mu(1-\varepsilon)}. \quad (36)$$

The first term in Eq. (35) represents the viscous energy loss, which is dominant in laminar flow where the Reynolds number is relatively low. The second term is the inertial energy loss, which becomes predominant in turbulent flow.

In our validation, a lattice packing of 8 spheres is considered as the porous media. The size of the domain equals that of the lattice packing, which is $2d_p \times 2d_p \times 2d_p$. Several different particle diameters are selected in the range $d_p$=10~80. The lattice packing is kept stationary and the flow is driven by a constant body force. Periodic boundary conditions are enforced for the fluid in all three directions. The fluid density, kinematic viscosity and relaxation parameter are 1,000 kg/m$^3$, $1\times10^{-4}$ m$^2$/s and 0.65, respectively. After the fluid develops into the steady state, the velocity at the inlet (or outlet) is measured to calculate the Reynolds number. The pressure drop through the packed bed is estimated from the pressure difference at two ends and the applied body force.

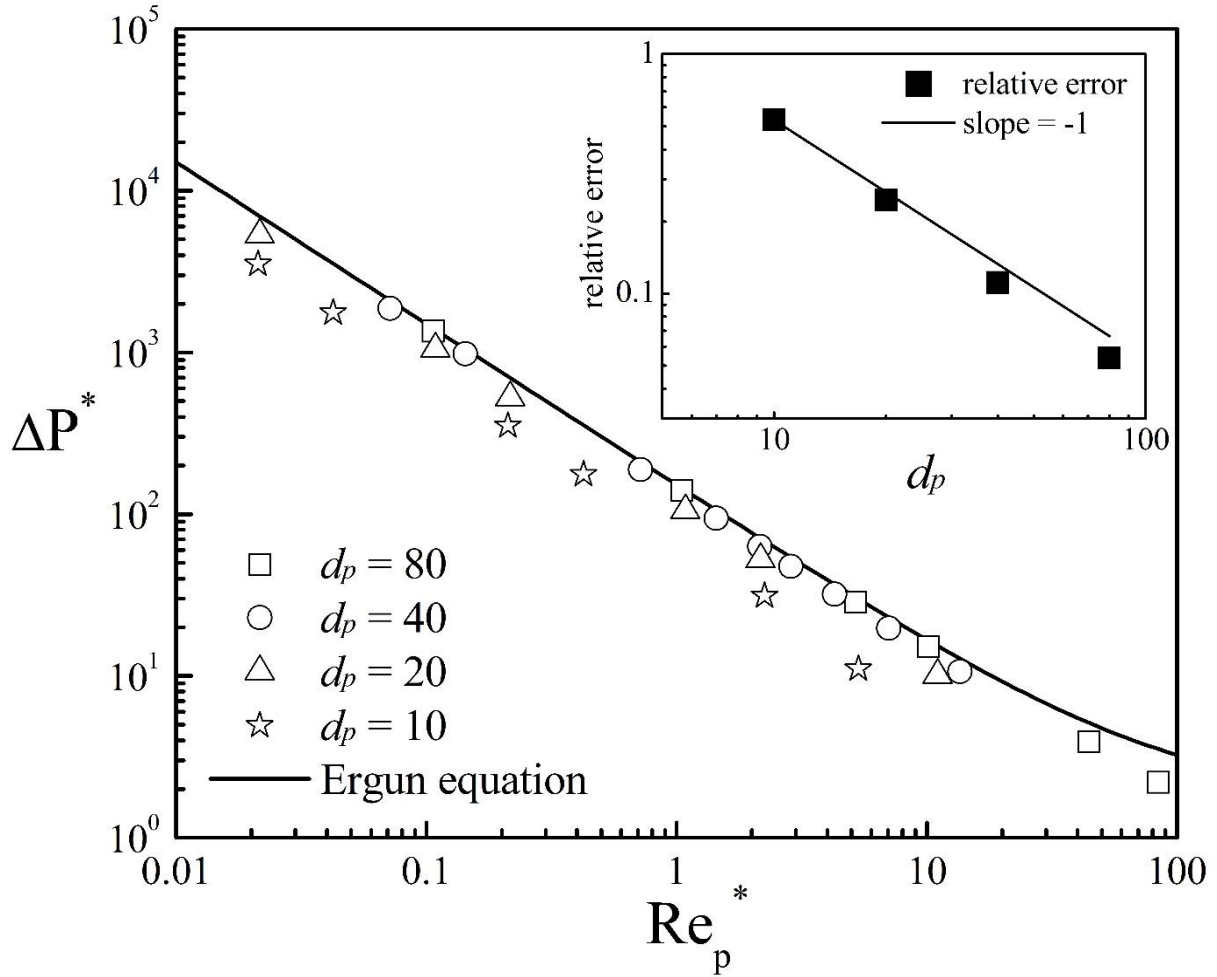

Fig. 11 The dimensionless pressure drop through a lattice packing as a function of the modified Reynolds number. The inset shows the relative error as a function of the lattice resolution.

Figure 11 shows the numerical results together with the prediction of Ergun equation. Generally, good agreement is observed within graphical accuracy when $Re_p^*$ is below 10, except for the case with $d_p$=10, where the lattice resolution is not sufficient. However, a large deviation is found when $Re_p^*$ is above 40. This is believed to be caused by the tuburlent effect, as can be inferred from Eq. (35). The average relative error of the dimensionless pressure drop is shown in the inset plot of Fig. 11, where the first order acuracy is obtained. Therefore, it can be concluded that our numerical method is able to well predict the flow through porous media.

## 3.5 Random packing of microspheres in quiescent fluid

Using the validated IBM-LBM-DEM model, random packing of microspheres in a quiescent fluid domain with and without particle adhesion is simulated. Initially, 500 monosized spheres are randomly placed in a cuboid box with size of 50×50×100, as shown in Figs. 12(a) and 12(e). Periodic boundary conditions are applied in both *x* and *y* directions, while no-slip wall boundaries are set on both top and bottom of the domain. The spheres settle under gravity with an initial velocity of (0.0, 0.0, -0.1) to accelerate the settling, which is in the same direction of the gravity. The diameter of the sphere is $4.8\times10^{-5}$ m, equivalent to 6 lattice units, and the mass density of the sphere is 3,000 kg/m$^3$. The surface energy of the sphere is fixed at 15 mJ/m$^2$ to account for the van der Waals adhesion. The fluid is quiescent at the beginning, with the density and kinematic viscosity of 1,000 kg/m$^3$ and $1\times10^{-6}$ m$^2$/s, respectively. The improved IBM method is employed in the simulation with the dimensionless relaxation parameter of 2. After a sufficient long time, all the kinetic energy of the spheres is dissipated and a mechanically stable packing is formed. A comparative simulation with the same configuration but without adhesion is also performed to further validate the adhesion model. Note that the non-adhesive Hertz contact model is used in the comparative simulation.

Figure 13 shows the snapshots of the packing process at different time instants for both adhesive and non-adhesive spheres. Generally, the packing formation process looks reasonable for both cases. The packing structure becomes stable after *t*=20,000. More importantly, we can see that the packing structure of the adhesive spheres is higher than that of the non-adhesive spheres, leading to a lower packing fraction, which agrees with the previous findings on random adhesive packings [42,61-62]. Then we quantify the representative properties, i.e. the global packing fraction and the mean coordination number to make a further comparison. The global packing fraction is defined as the ratio of the total volume of the spheres to the total volume that the packing occupies. The coordination number is defined as the number of the neighbouring particles that are in contact with a reference sphere, which is calculated by judging whether the centre-to-centre distance of two spheres is smaller than the sum of their radii. The mean coordination number is then obtained through the average over all the spheres.

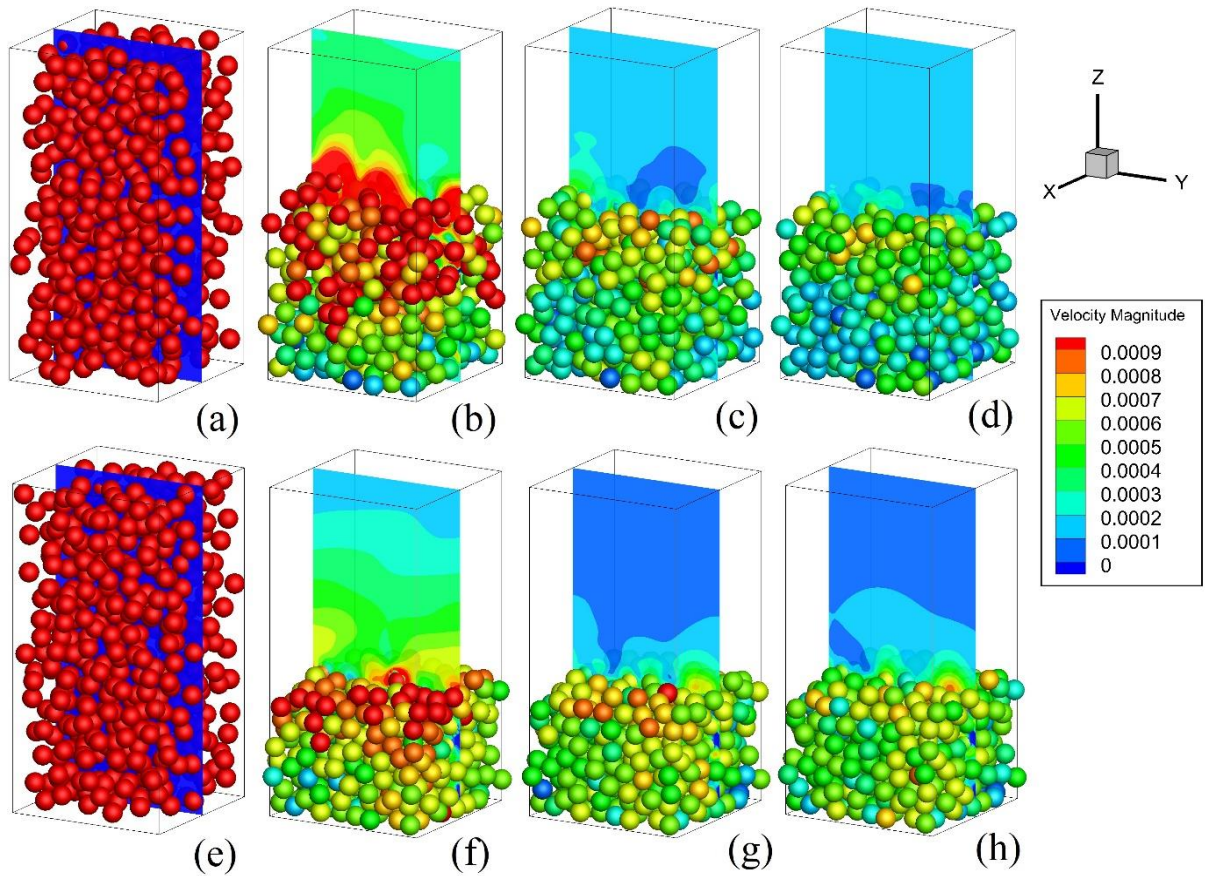

Fig. 12 The random packing of 500 spheres in quiescent fluid at (a)(e) $t$=0, (b)(f) $t$=5,000, (c)(g) $t$=20,000, (d)(h) $t$=100,000. (a) (b) (c) (d) denote the adhesive spheres, while (e) (f) (g)

(h) stand for the non-adhesive spheres. The contour plot shows the velocity magnitude of the flow field in the slice of $x$=25 in the $x$-plane.

Figure 13(a) presents the time evolution of the global packing fraction $\phi$ and the mean coordination number $Z$ for both adhesive and non-adhesive spheres. It can be seen that that both $\phi$ and $Z$ rise quickly from $t$=0 to $t$=20,000 in lattice unit, during which the particles are still in settling and start to form contact network. After $t$=20,000, the packing structure enters the relaxation stage and gradually becomes stable. However, obvious distinctions can be observed between adhesive and non-adhesive spheres. The global packing fraction and the mean coordination number of adhesive spheres are $\phi=0.495$ and $Z$=5.08, which are lower than those of non-adhesive spheres, $\phi=0.557$ and $Z$=5.50. The result for non-adhesive spheres is in quantitative agreement with the random loose packing (RLP) limit of granular matter [63], while the result for adhesive sphere is below the RLP limit, agreeing with the previous investigations [42,64]. Further comparison can be found in Fig. 13(b), showing the probability distribution function $P(z)$ of the local coordination number of each sphere. For non-adhesive spheres, most of the spheres have 6 neighbours, while for adhesive spheres, the majority of the spheres tends to have only 4 or 5 contacts. This is because adhesion can provide additional resistances to prevent the sphere from rolling over other spheres, thus fewer contacts are needed to reach a local mechanical equilibrium [61].

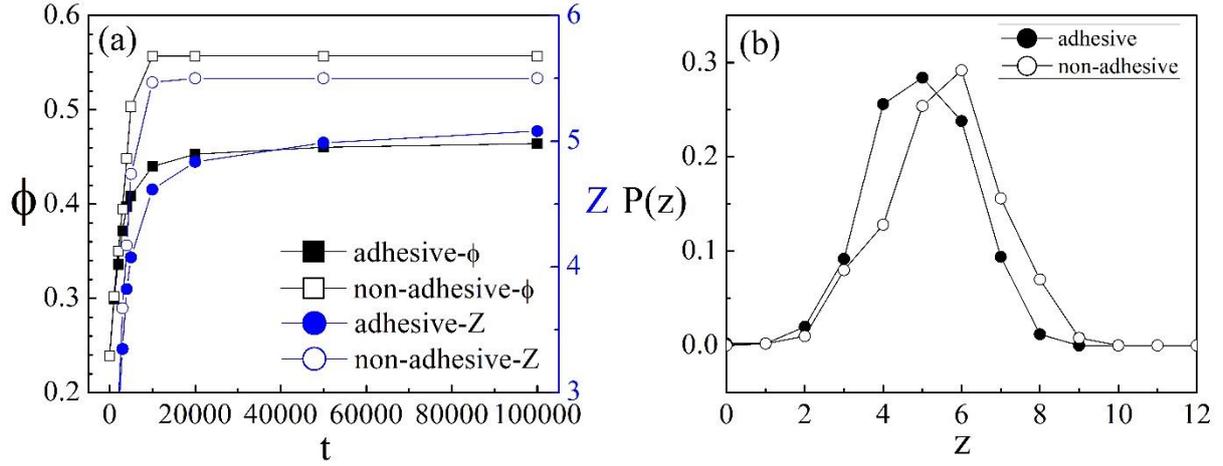

Fig. 13 (a) The global packing fraction and the mean coordination number as a function of the computational time for both adhesive and non-adhesive spheres. (b) The probability distribution function of the local coordination number.

The results of adhesive sphere are further compared with those from simulations and experiments of wet granules [65], as displayed in Table 2. In both Refs. [65] and [66], the results were obtained with particles of 250 μm and a liquid content of 20%, which is distinguished from the setup of our simulation. It should be noted that there were very few studies on the random packing of adhesive spheres immersed in liquid. Therefore the random packing of wet granules with large liquid content is only alternative we could choose to compare. From Table 2, it is observed that our simulation is in qualitative agreement with the literature. As a consequence, it is concluded that the adhesive contact model implemented in the hybrid IBM-LBM-DEM framework is capable of capturing the adhesive mechanics between microspheres effectively.

Table 2 Comparison of the global packing fraction and mean coordination number for

|  | adhesive spheres | | |
| --- | --- | --- | --- |
| Parameters | Present work | Ref. [65]-Exp | Ref. [66]-Sim |
| $\phi$ | 0.495 | 0.474 | 0.445 |
| $Z$ | 5.08 | Not available | 5.70 |

# 4. Conclusions

In this paper, an efficient LBM-DEM numerical framework for modelling adhesive particle-liquid flow is developed. An improved immersed boundary method, which accurately describes the multi-intersection of more than one particle with the same lattice cell, is implemented for solving the fluid-particle interactions. A linear approximation method is developed to determine the solid fraction in a local lattice cell, which is validated to be of high efficiency and accuracy compared with other approaches. In the DEM, the JKR adhesive contact model is adopted to account for the particle-particle normal force. Other dissipative interactions, including the sliding, twisting and rolling resistances, are all well considered with a spring-dashpot-slider model in the presence of adhesion. Validated with several benchmark cases, the hybrid IBM-LBM-DEM numerical approach is proved to be capable of providing more detailed and accurate results in the computation of fluid flow between dense particle arrays, as well as capturing the adhesion between microspheres. An application of the numerical approach to the random packings of adhesive microspheres is also performed. The results show that the packing density of non-adhesive microspheres in quiescent fluid agrees with the RLP reported in previous experiments but with a slightly higher mean coordination number. With the introduction of van der Waals adhesion, the packing properties go below the RLP limit. However, considering both the small size scale and short collision time scale of micro-sized particles, this numerical framework might not be suitable for simulations across a large length scale. Further improvements in efficiency can be realised with parallel computing on GPUs or other high performance computing clusters.

# Acknowledgement

This work is funded by the Engineering and Physical Sciences Research Council, UK (EPSRC, Grants No: EP/N033876/1). The authors are grateful to Dr. Duo Zhang and Dr. Nicolin Govender at the University of Surrey, Prof. Shuiqing Li at Tsinghua University and Dr. Sheng Chen at Huazhong University of Science and Technology for their helpful suggestions and fruitful discussion in developing the LBM-DEM numerical approach.